\documentclass[pra,twocolumn,showpacs,superscriptaddress]{revtex4-1}
\usepackage{graphicx}
\usepackage{amssymb,amsmath}
\usepackage{epsfig}
\usepackage{txfonts}

\usepackage[unicode]{hyperref}
\hypersetup{
    pdftitle={Clustering},
    pdfauthor={Yu, Billam, Nian, Reeves, Bradley},
    pdfsubject={Bright Soliton Arrest},
    colorlinks=true,
    linkcolor=blue,
    citecolor=blue,
    filecolor=black,
    urlcolor=blue
}
\def\urlprefix{}
   \def\url#1{}

\newcommand{\eref}[1]{Eq.~\eqref{#1}}

\newcommand{\be}{\begin{equation}}
\newcommand{\ee}{\end{equation}}
\newcommand{\bea}[1]{\begin{align}#1\end{align}}

\newcommand{\nn}{\nonumber }

\begin{document}
\title{Theory of the vortex-clustering transition in a confined
    two-dimensional quantum fluid}

\author{Xiaoquan Yu}
\email{xiaoquan.yu@otago.ac.nz}
\affiliation{Department of Physics, Centre for Quantum Science, and Dodd-Walls Centre for Photonic and Quantum Technologies, University of Otago, Dunedin, New Zealand. } 
\author{Thomas P. Billam}
\affiliation{Joint Quantum Centre (JQC) Durham-Newcastle, School of Mathematics and Statistics, Newcastle University, Newcastle upon Tyne, NE1 7RU, United Kingdom.}
\affiliation{Joint Quantum Centre (JQC) Durham-Newcastle, Physics Department, Durham University, Durham, DH1 3LE, United Kingdom.}
\author{Jun Nian}
\affiliation{Institut des Hautes \'Etudes Scientifiques, Bures-sur-Yvette, France.}
\affiliation{C. N. Yang Institute for Theoretical Physics, Stony Brook University, Stony Brook, U.S.A.}
\author{Matthew T. Reeves}
\affiliation{Department of Physics, Centre for Quantum Science, and Dodd-Walls Centre for Photonic and Quantum Technologies, University of Otago, Dunedin, New Zealand. } 
\author{Ashton S. Bradley}
\email{ashton.bradley@otago.ac.nz}
\affiliation{Department of Physics, Centre for Quantum Science, and Dodd-Walls Centre for Photonic and Quantum Technologies, University of Otago, Dunedin, New Zealand. } 
\date{\today}

\begin{abstract}
Clustering of like-sign vortices in a planar bounded domain is known to occur at negative temperature, a phenomenon that Onsager demonstrated to be a consequence of bounded phase space. In a confined superfluid, quantized vortices can support such an ordered phase, provided they evolve as an almost isolated subsystem containing sufficient energy. A detailed theoretical understanding of the statistical mechanics of such states thus requires a microcanonical approach.
Here we develop an analytical theory of the vortex clustering transition in a neutral system of quantum vortices confined to a two-dimensional disc geometry, within the microcanonical ensemble.  The choice of ensemble is essential for identifying the correct thermodynamic limit of the system, enabling a rigorous description of clustering in the language of critical phenomena. As the system energy increases above a critical value, the system develops global order via the emergence of a macroscopic dipole structure from the homogeneous phase of vortices, spontaneously breaking the $\mathbb Z_2$ symmetry associated with invariance under vortex circulation exchange, and the rotational $\rm SO(2)$ symmetry due to the disc geometry. The dipole structure emerges characterized by the continuous growth of the macroscopic dipole moment which serves as a global order parameter, resembling a continuous phase transition. The critical temperature of the transition, and the critical exponent associated with the dipole moment, are obtained exactly within mean field theory. The clustering transition is shown to be distinct from the final state reached at high energy, known as supercondensation. The dipole moment develops via two macroscopic vortex clusters and the cluster locations are found analytically, both near the clustering transition, and in the supercondensation limit. The microcanonical theory shows excellent agreement with Monte Carlo simulations, and signatures of the transition are apparent even for a modest system of 100 vortices, accessible in current Bose-Einstein condensate experiments.
\end{abstract}

\maketitle
\section{Introduction}
In two-dimensional (2D) classical fluids, giant coherent vortex structures can emerge from microscopic vortex excitations~\cite{Eyi2006.RMP78.87,SIGGIA:1981up,Boffetta12a}, as end states of turbulent fluid dynamics involving an inverse energy cascade~\cite{Kraichnan:1975ku,KRAICHNAN:1980uy}. 
The Great Red Spot in Jupiter's atmosphere~\cite{Turkington:2001jx,Bouchet:2002je,Choi:2011ff} is a well-known example. Onsager explained this spontaneous formation of large-scale vortices from an equilibrium statistical mechanics 
point of view by studying a point-vortex model in a bounded domain~\cite{Ons1949.NC6s2.279,Eyi2006.RMP78.87}. The phenomenon stems from the bounded phase space, which supports negative temperature
states that favour the spontaneous clustering of like-sign vortices. 
As an equilibrium statistical mechanics problem, cluster formation in 2D vortex systems has attracted much attention~\cite{Joyce:OIPPoHnh,Montgomery:1974dr,Edwards:1974cs, Williamson:1977ua,Book:1975dd,Pointin:1976gf,KRAICHNAN:1980uy,Campbell:1991wt,Smith:1989a,Smith:1990cx,Yatsuyanagi:2005jm,Esler:2015ji}. Building on Joyce and Montgomery's formulation of vortex statistical mechanics~\cite{Joyce:OIPPoHnh}, 
Smith and O'Neil~\cite{Smith:1989a,Smith:1990cx} investigated the single charge 2D plasma using the point-vortex model in a rotating container and showed that the formation of a non-axisymmetric cluster resembles a second-order phase transition. The analogy to a phase transition is a powerful tool and provides deep insights into the nature of vortex cluster formation.    
To the best of our knowledge, the nature of the transition to large-scale clustered states in bounded {\em neutral} vortex systems remains an open problem 
and fundamental questions remain unanswered: What is the exact transition temperature, and the universality class of the transition? What is the precise mechanism of cluster formation? 
A rigorous theory capable of addressing these questions may expand our knowledge of critical phenomena~\cite{Cardy:1996a,*Goldenfeld:1992a,*Kardar:2007um} and pattern formation~\cite{Cross:1993el}, and lead to a deeper understanding of collective vortex states in a broad class of 2D systems~\cite{Chorin:1994ig,*Pismen1999}.   \par

In this work we develop an analytical treatment within mean field theory~\cite{Joyce:OIPPoHnh,Smith:1990cx} of the clustering transition in a neutral point-vortex model confined to a uniform disc geometry. The theory provides a systematic description of the negative temperature phenomena in terms of inverse temperature $\beta(\varepsilon)$, a function of the system energy $\varepsilon$ parametrizing microcanonical states. 
We analytically find $\beta_c<0$ at which a clustering transition occurs, whereby large-scale aggregation of like-sign vortices can occur, breaking the underlying symmetries of the system.  Our mean field analysis gives analytical expressions for the structure of the clustered phase and critical behavior of the macroscopic dipole moment that serves as a global order parameter. In contrast to conventional second-order phase transitions linked to a local order parameter, the vortex clustering transition thus develops via the emergence of global order characterized by a non-zero macroscopic dipole moment. At high energy the system enters the supercondensation limit at $\beta_s<\beta_{c}$~\cite{Kraichnan:1975ku}, involving point-like concentration of vortices with the same circulation, and divergence of the system energy.
The supercondensation does not break any symmetry (it occurs within the symmetry broken phase) and is an end
point of a continuous process as energy increases. We test the microcanonical theory against Monte Carlo (MC) sampling of the microcanonical ensemble for finite vortex numbers, and find good agreement even for a modest system containing a total of $N=100$ vortices. To probe the role of finite vortex number further, we compute the energy dependence of the macroscopic dipole moment, for a range of $N$ from 50 to 1000. The critical exponent determining the order parameter scaling emerges clearly even for $N=100$ vortices.

\section{Background}
\subsection{Negative Temperature Vortex States}

As a measure of motion, absolute temperature is always positive. However under certain conditions, negative temperature states can occur for an almost isolated subsystem. The concept of temperature is applicable to such a subsystem if it is in a long-lived metastable state.
Since the coordinates $(x,y)$ of
a vortex are conjugate variables, 
the crucial feature of 2D point-vortex systems in a bounded domain is that the phase space
volume is finite. Let the phase space volume be $\Omega (\varepsilon)$ for a given energy $\varepsilon$, then the thermodynamic entropy is $k_B \ln \Omega(\varepsilon)$, where $k_B$ is the Boltzmann constant. For the extreme situation in which all the vortex dipoles (vortex-antivortex pairs) collapse, we then have $\Omega(\varepsilon=-\infty)=0$. The opposite limit occurs when all the like-sign vortices concentrate into a point, in which case $\Omega(\varepsilon=\infty)=0$. It is clear that $\Omega(\varepsilon)$ must reach its maximum at an intermediate energy $-\infty<\varepsilon_m<\infty$. Then   
the inverse temperature
\bea{\frac{1}{T}\equiv k_B \frac{\partial \ln \Omega(\varepsilon)}{\partial \varepsilon}} is thus negative for $\varepsilon>\varepsilon_m$.  High energy equilibrium vortex states at negative temperature correspond to large-scale clusters~\cite{Ons1949.NC6s2.279}.
Negative temperature states can arise in systems with finite and discrete degrees of freedom, for instance localized spin systems~\cite{Purcell:1951hi,Oja:1997ci,Medley:2011jo}. Recently such negative temperature states have been realized for motional degrees of freedom~\cite{Braun:2013gw}, where the finite band width sets an upper bound for the energy spectrum~\cite{Carr:2013kv}.

\subsection{Motivation from Bose-Einstein condensates}
An interesting scenario for potential realization of Onsager vortex states may occur in planar Bose-Einstein condensates (BECs). In atomic BECs the point-vortex model provides an excellent description of vortex dynamics in a hydrodynamic regime~\cite{Fetter:1966uo,Rooney:2011fm,Aioi2011,Moon:2015hn}. Furthermore, experimental observations of small-scale clustering~\cite{Neely:2010gl,Neely:2013ef}, and Gross-Pitaevskii simulations demonstrating clustering in forced~\cite{Bradley:2012ih,Reeves:2013hy} and decaying \cite{Billam:2014hc} homogeneous systems, in harmonic traps~\cite{White:2012vz}, and disc geometries~\cite{Simula:2014ku} have raised the prospect that large-scale clustered states may be realized, amid increasing theoretical~\cite{Reeves:2014a,Stagg:2014hl,Lucas:2014kf,Powis:2014hy,Lucas:2014kf,Billam:2015fj,Stagg:2015vt,Reeves:2015dt,Skaugen:2016ct,Skaugen:2016hq,Salman:2016vt} and experimental~\cite{Kwon:2014ud,Wilson:2015cj,Kwon:2015hh,Moon:2015hn,Kim:2016tm} interest.
\par
The scenario of negative temperature vortex states within a low positive temperature superfluid may appear implausible as the superfluid in a BEC is compressible and the vortex-sound coupling must be taken into account. In general, vortex-sound coupling can play an important role in a compressible superfluid~\cite{Pismen1999,Parker04a,Parker:2012je}, however, there are also regimes where the acoustic loss rate can be much slower than the dynamical timescales of vortex evolution. Several theoretical works have considered the viability of negative temperature vortex states using the Gross-Pitaevskii equation (GPE)~\cite{Billam:2014hc,Simula:2014ku,Reeves:2014a}. They show that the compressibility of the quantum fluid does not prevent the clustering of like-sign vortices, and moreover it can even provide a mechanism to drive clustering, via an evaporative heating process~\cite{Simula:2014ku}, whereby annihilations of low-energy vortex-antivortex pairs ``heat up" the remaining vortices and the end state shows a clustered configuration of like-sign vortices. Experimental observations that like-sign vortices can aggregate into long-lived multiply charged structures provides further evidence that such a regime may be accessible~\cite{Neely:2010gl,Neely:2013ef}. These experiments can be seen as first steps towards realizing large-scale, negative temperature, clustered states of quantum vortices.  
\par
Several works have considered the possibility of an inverse energy cascade in two-dimensional BEC, suggesting that the compressibility of the fluid may be fatal for large-scale clustering, instead causing transport of energy to small scales driven by vortex dipole annihilation~\cite{Numasato:2010ba,Numasato:2010ir,Chesler:2013bl}. However, these works considered specific scenarios that immediately prohibited clustering, due to either starting from initial conditions dominated by acoustic energy~\cite{Numasato:2010ba,Numasato:2010ir}, or evolving the system according to an over-damped equation of motion~\cite{Chesler:2013bl}. As shown in a systematic study of energy transport~\cite{Billam:2015fj}, a clear regime exists where energy is transported to large scales via a vortex clustering process. The regime is one of weak dissipation, and large vortex number. Recent analytical calculations further support the possibility that the  sound-vortex interaction is not strong enough to prevent energy transport to large scales~\cite{Lucas:2014kf}, consistent with GPE simulations~\cite{Bradley:2012ih,White:2012vz,Reeves:2013hy,Billam:2014hc,Simula:2014ku,Reeves:2015dt}. 
\par
In summary, although the temperature of the bulk liquid is always positive, rendering the negative temperature states ultimately unstable, if the temperature of a BEC is low enough and the mean distance between vortices is much larger than the healing length, the clustered phase is expected to be long-lived. Indeed, GPE simulations indicate that a regime exists where a separation of timescales allows clustered states to be long-lived. The cause of this timescale separation appears to have two sources: (i) the vortex-antivortex recombination rate is rather slow in 2D (stemming from weak coupling to the sound field; this weak coupling also holds for single vortices), (ii) clustering of same sign vortices suppresses close approaches of vortices and antivortices. Thus, in a long time window, a system of 2D quantum vortices may evolve as an almost isolated subsystem of a confined BEC at sufficiently low temperature. Such quasi-equilibrium states may also have connections with non-thermal fixed point behavior~\cite{Nowak:2013tl,Langen:2016bu}. While much evidence points to a weak coupling regime, a detailed microscopic picture of the mechanism remains an open problem and such states have yet to be observed in the laboratory, posing a exciting future challenge for both theoretical and experimental work.

\subsection{Point-Vortex Regime}
We consider a total of $N$ quantum vortices in a planar superfluid, with core size determined by the healing length $\xi$.
 In a BEC at sufficiently low temperature, coupling to the thermal cloud is negligible, and provided that the average distance between vortices greatly exceeds the healing length, vortices are only weakly coupled to phonon modes.
The point-vortex model then applies, and the other degrees of freedom can be neglected. In this regime the point-vortex model can also be rigorously obtained from the GPE~\cite{Fetter:1966uo,Billam:2014hc} and the incompressible kinetic energy spectrum of quantum vortices in the GPE is well approximated by the point-vortex energy spectrum~\cite{Bradley:2012ih,Billam:2014hc}. 

Let us now consider $N=N_++N_-$ quantum vortices consisting of $N_+$ vortices and $N_-=N_+$ anti-vortices confined to a uniform disc with radius $R$, by a hard-wall boundary.  Hereafter we set $R=1$, giving the point-vortex Hamiltonian~\cite{Hess:1967ke,Newton:2001} 
\bea{
\label{Hamiltonian1}
 H &= -\sum_{i\neq j} \kappa_i \kappa_j \ln|{\mathbf r}_i-{\mathbf r}_j|+ \sum_{i,j}\kappa_i \kappa_j \ln\bigg|({\mathbf r}_i-\bar{\mathbf{r}}_j)|\mathbf{r}_j|\bigg|,
}
where $\kappa_{i}=\pm 1$ is the circulation of 
a vortex at position ${\mathbf r}_i$. 
The image terms are given by the second summation in \eref{Hamiltonian1} and 
$\bar{\mathbf{r}}_j={\mathbf r}_j/|{\mathbf r}_j|^2$ 
is the position of the $j$th image vortex. The image terms ensure that the velocity normal to the boundary vanishes; without losing generality we choose images such that the stream function is zero on the boundary. The Hamiltonian \eqref{Hamiltonian1} has $\mathbb Z_2$ symmetry (invariant under $\kappa_i \rightarrow -\kappa_i$) and rotational $\rm SO(2)$ symmetry due to the disc geometry. In a dilute gas BEC the Hamiltonian Eq.~\eqref{Hamiltonian1} is measured in the energy unit $E_0=\rho m_a \kappa^2/4\pi$, 
where $\rho$ is the superfluid density, $\kappa=h/m_a$ is the circulation quantum and $m_a$ is the atomic mass.  
\eref{Hamiltonian1} describes well-separated ($\xi \sqrt{N}\ll 1$) quantum vortices in 2D incompressible superfluids.
Note that the second summation of Eq.~\eqref{Hamiltonian1} depends on the shape of the domain and is absent for an unbounded domain. 
The disc geometry is chosen not only because it can be realized in current BEC experiments~\cite{Henderson:2009eo,Gaunt:2013ip,Gauthier:2016vs} but also the simple form of the domain leads to the possibility of an exact analytical treatment. 
In general the point-vortex model also describes vortices in classical inviscid, incompressible fluids~\cite{Aref:1999gt,Pointin:1976gf,KRAICHNAN:1980uy,Campbell:1991wt}, and guiding-center plasma dynamics~\cite{Joyce:OIPPoHnh,Edwards:1974cs}.

\subsection{Scaling and Continuous Hamiltonian}
The model Eq.~\eqref{Hamiltonian1} exhibits negative temperature states at high energies. In the standard thermodynamic limit, namely $N \rightarrow \infty$  and $V \rightarrow \infty$ with $H/N$ and $N/V$ held fixed, where $V$ is the area, negative temperature states do not exist. In order to support negative temperature states, a bounded domain is necessary. A well-defined thermodynamic limit for the vortex system in a bounded domain is found by a careful choice of scaling~\cite{Eyi2006.RMP78.87,Eyink:1993vw,Caglioti:1995tc}. In the clustered phase, the energy $H\sim N^2$ is due to the sum over all vortex pairs. Hence $ H/N^2$ is finite-valued as $N \rightarrow \infty$. Moreover the entropy per vortex is bounded by the area of the domain. A formulation of the theory with a well-defined thermodynamic limit that can describe negative temperature states is thus given by the rescaled Hamiltonian 
\bea{\label{scH}
{\cal H}&\equiv 4H/N^2,
}
equivalent to introducing the rescaling $\kappa_{i} \rightarrow \kappa_{i}/N_{\pm}=2\kappa_{i}/N$.
 The validity of this rescaling for the bounded vortex system has been verified rigorously~\cite{Eyink:1993vw,Caglioti:1995tc}.
\par
A course-grained formulation of \eref{scH} valid for $N\gg 1$ has been derived~\cite{Joyce:OIPPoHnh}, and we take the same approach here. Defining the local density of positive (negative) vortices
\bea{\label{ni}
n_{\pm}(\mathbf {r})\equiv  \frac{1}{N_{\pm}}\sum_{i}\delta(\mathbf{r}-\mathbf{r}^{\pm}_{i}),
}
satisfying the normalization condition $\int d^2 \mathbf {r}\;  n_{\pm} =1$, 
where $\mathbf{r}^{\pm}_i$ is the position of the vortex $i$ of circulation $\pm \kappa_{i}$ respectively,
then the vorticity field is
\bea{\label{sigd}
\sigma (\mathbf{r})\equiv n_+(\mathbf{r})-n_-(\mathbf{r}).
}
At large $N_{\pm}$, the continuous coarse-grained Hamiltonian reads
\bea{
\label{Hamiltonian2}
{\cal H}_{\rm eff}&=\frac{1}{2} \int d^2 {\mathbf r}\; d^2{\mathbf r}' \;\sigma({\mathbf r})\phi({{\mathbf r},{\mathbf r}'})\sigma({\mathbf r}'),
}
which gives a well-defined continuum limit of Eq.~\eqref{Hamiltonian1} as $N\to\infty$ and
in the clustered phase ${\cal H}_{\rm eff} \sim \mathcal{O}(1)$~\cite{Eyi2006.RMP78.87,Eyink:1993vw,Caglioti:1995tc}. 
The potential $\phi(\mathbf{r},\mathbf{r}')$ is the Green's function of the Laplace operator
\bea{
\nabla^2 \phi(\mathbf{r} ,\mathbf{r}')=-4\pi \delta (\mathbf{r} -\mathbf{r}'),
}
where $\phi (\mathbf {r},\mathbf {r}')=\phi(\mathbf {r}',\mathbf {r})$ , and $\phi (\protect \mathbf {r},\protect \mathbf
	{r}')\sim -2 \protect \qopname \relax o{log}|\protect \mathbf {r}-\protect
	\mathbf {r}'|$ as $|\protect \mathbf {r}-\protect \mathbf {r}'|\rightarrow
	0$~\cite {Lin:LboE0Oun}.
The stream function \bea{\psi(\mathbf{r})\equiv\int d^2 \mathbf{r}'\phi(\mathbf{r},\mathbf{r}')\sigma (\mathbf{r}')}
satisfies the Poisson equation 
\bea{
\label{Poisson}
\nabla^2 \psi = - 4 \pi \sigma(\mathbf {r}),
}
with the boundary condition $\psi(r=1,\theta)=C$. Here $\theta$ is the azimuthal angle and $C$ is a constant chosen to be zero, which is equivalent to including image terms in \eref{Hamiltonian1}. Recall that the radial velocity $ v_r =r^{-1} \partial \psi
/\partial \theta $, and this boundary condition ensures that there is no flow
across the boundary of the container.

In the disc container, the energy 
\bea{
{\cal H}_{\rm eff}=E=\frac{1}{2} \int d^2 {\mathbf r}\;\sigma \psi = \frac{1}{8\pi}\int d^2 {\mathbf r}\; |\nabla\psi|^2 \geq 0}
and the angular momentum per vortex
\bea{
L&=\int d^2 \mathbf{r}\; r^2\sigma} 
are conserved quantities.
\subsection{Statistical Mechanics}
Let us briefly recall the statistical mechanics of the system developed by Joyce and Montgomery~\cite{Joyce:OIPPoHnh}, who derived self-consistent equations for the vorticity field via a maximum entropy principle. 
	\par
	The most probable density distribution of vortices is obtained by maximizing the entropy per vortex
\bea{
\label{entropy}	
S= - \int d^2 \mathbf{r} \left(n_+ \ln n_+ +n_- \ln n_-\right),
}
with $N_+$, $N_-$, $E$ and $L$ constrained to fixed values in the microcanonical ensemble. The variational equation 
\bea{  
\delta S -\beta \delta E- \alpha \delta L- \mu_+ \delta N_+/N_+ -\mu_- \delta N_-/N_- =0,}
where $\beta$, $\alpha$ and $\mu_\pm$ are Lagrange multipliers, gives the maximum entropy state
\bea{
\label{selfconsistenteq}
n_\pm (\mathbf{r}) &= \exp \left[\mp \beta \psi(\mathbf{r})\mp \alpha r^2 +\gamma_\pm \right],
}
with $\gamma_{\pm}=-\mu_{\pm}-1$. Here $\beta=\beta(E)$ is the dimensionless inverse temperature describing the vortex subsystem at fixed energy, $\alpha$ is proportional to the rotation frequency of the container, and $\gamma_\pm$ set the normalization. Eq.~\eqref{selfconsistenteq} together with Eq.~\eqref{Poisson} provides a mean-field description of the distribution of vortices in a disc~\cite{Eyi2006.RMP78.87}. 
\section{Clustered phases}
In the following we investigate possible stable large-scale coherent structures described by Eqs.~\eqref{Poisson},~\eqref{selfconsistenteq}.
Non-trivial solutions for the vortex density exist only for
$\beta <0$. We note that some exact solutions of Eqs.~\eqref {Poisson},~\eqref {selfconsistenteq}
in rectangular domains are known~\cite
{TING:1984je,TING:1987cu,Kuvshinov:2000fn,Gurarie:2004dz}, however, we are unaware of any exact solutions for the disc geometry. Our approach is to extend the bifurcation theory of Smith and O'Neil~\cite{Smith:1989a,Smith:1990cx}, originally developed by  for the single charge vortex system, to the neutral vortex system confined to a disc.

\subsection{Constrained Variation of the Maximum Entropy State}
Let us start at a solution $n_{\pm}$ of Eqs.~\eqref{Poisson},~\eqref{selfconsistenteq} at $E$ and $L$, and consider a nearby 
solution $n_\pm + \delta n_\pm$ at $E+\delta E$ and $L+\delta L$. 
The corresponding changes of the constraints are
\bea{ \label{constraint1}
0&=\int d^2 {\mathbf r} \;\delta n_{\pm}, \\
\label{constraint2}
\delta E&= \int d^2 \mathbf{r}\; \psi \delta \sigma + \frac{1}{2}\int d^2 \mathbf{r}\; \delta\psi \delta \sigma, \\
\label{constraint3}
\delta L&= \int d^2  \mathbf{r}\; r^2 \delta \sigma.} 
The variation of \eref{Poisson} is
\bea{\label{VP}
	(4 \pi )^{-1}\nabla^2 \delta \psi & = -  \left[ \delta n_+ (\mathbf{r}) - \delta n_- (\mathbf{r})\right] \nn\\
 &= n(\psi \delta \beta + \beta \delta \psi+  r^2 \delta \alpha) - n_+ \delta \gamma_+ +  n_- \delta \gamma_-,} 
where $n=n_+ + n_-$, and $\delta \gamma_{\pm},\delta\beta$ and $\delta \alpha$ can be 
expressed as linear functions of $\delta E$, $\delta L$ and $\delta \psi$.
In contrast to the single charge vortex system~\cite{Smith:1989a,Smith:1990cx}, the macroscopic dipole formation in the neutral vortex system does not require finite rotation frequency of the container, and energy (equivalently temperature) is the only control parameter of the transition for given angular momentum $L$.
\par
We now focus on the non-rotating container, setting
$\alpha=0$. Then the homogeneous state $n_\pm=n_0/2=1/\pi$ is the solution of Eqs.~\eqref{Poisson},~\eqref{selfconsistenteq}, at which $\sigma=0$, $\psi=0$, $E=0$ and $L=0$.
Evaluating the variations of the constraints Eqs.\eqref{constraint1}, \eqref{constraint2}, \eqref{constraint3} from the homogeneous state to the leading order in $\delta \psi$ gives 
\bea{\label{condition1}
\delta \gamma_+ +\delta \gamma_-&=0, \\
\label{condition2}
\delta E&=0, \\
\label{condition3}
\delta \gamma_+&=\frac{\beta n_0}{2}\int d^2 \mathbf{r}\; \delta \psi.}
Plugging Eqs.~\eqref{condition1}, \eqref{condition2}, \eqref{condition3} into Eq.~\eqref{VP}, we obtain the zero mode ($\delta E=0$) equation 
\bea{
\label{zeromode}
\left[\nabla^2 - 4 \pi \beta n_0 (1 - \hat {\cal I})\right] \delta \psi = 0,
}
where the integral operator is defined as
\bea{\hat{\cal I}\delta \psi\equiv\frac{1}{\pi}\int d^2\mathbf{r}\; \delta \psi,
} 
and the solutions are required to satisfy the boundary condition $\delta \psi(r=1,\theta)=0$.
The onset of clustering occurs if \eref{zeromode} has non-zero solutions for the fluctuation $\delta\psi$. Note that in our mean field approach the value of $\beta $ is undefined in the homogeneous phase, and its precise value at the transition is determined by the specific unstable mode that develops at $E=0$. 

\subsection{Vortex Clustering Transition}
We now give a formulation of the clustering transition as a symmetry-breaking perturbation of the homogeneous neutral vortex configuration, and investigate the system dependence on energy near the transition.
It is natural to decompose \eref{zeromode} into azimuthal Fourier harmonics 
\bea{\psi_{l}(r,\theta) = \hat{\psi}_{l}(r) e^{i l \theta}} with the integer mode number $l$.
The normalization condition for $\psi_l$ is chosen to be 
\bea{\label{normalization}
-\int d^2 \mathbf{r}\; \psi_l \nabla^2 \psi_l&=4\pi
}
for convenience. 
Let us now consider a small variation along the mode $l$, $\delta \psi=\delta \psi_l \equiv \epsilon \psi_l$, where $\epsilon$ is a small amplitude.

In this work we focus on $l>0$ modes, which break the spatial $\rm SO(2)$ symmetry as well as the $\mathbb Z_2$ symmetry. Breaking the $\mathbb Z_2$ symmetry while preserving the $\rm SO(2)$ symmetry involves the $l=0$ mode, and non-zero $L$. This polar angle $\theta$-independent clustered phase is not pursued further here.
For $l>0$, $\hat{\cal I}\delta \psi=0$, and consequently $\delta \gamma_{+}=\delta \gamma_{-}=0$ and $\delta L=0$.
Then \eref{zeromode} is an eigenvalue problem of a Laplacian operator: 
\bea{\label{smode}
\left[\frac{1}{r} \frac{d}{dr} r \frac{d}{dr} - \frac{l^2}{r^2} \right] \psi_{l} &= - \lambda \psi_{l},
}
with the Dirichlet boundary condition ${\psi}_{l}(r=1,\theta)=0$. \eref{smode} has non-zero solutions only when $\lambda=j^2_{l,m}$ , where $j_{l,m}$ is the 
$m$th positive zero of the Bessel function of the first kind $J_l(r)$. Due to the dipolar form of the dominant symmetry breaking perturbation, the root $j_{1,1}\simeq 3.832$ plays an important role in what follows.
The onset of the $\theta$-dependent clustered stable phase occurs when the
$l=1$ and $m=1$ mode starts to emerge at
\bea{\label{bc}
\beta_c&=-j_{1,1}^2/4 \pi n_0 \simeq-1.835,
}
and the corresponding normalized non-zero solution is 
\bea{\label{dipMode}
\psi_1(r,\theta)&=2\sqrt{2}\left[|J_0(j_{1,1})|j_{1,1}\right]^{-1} J_1(j_{1,1}r)\cos(\theta).
}
Here we choose the positive $y$-axis as the dipole direction without loss of generality. 
Higher values of $l$ and $m$ correspond to additional ordered phases with lower entropy 
at finite energy, and hence have less statistical weight.
For zero angular momentum $L=0$ and given energy $E \neq 0$, 
the maximum entropy state arises from $l=1$ and $m=1$ mode.
We refer to this $\rm SO(2)$ symmetry breaking phase as the dipole phase.  

Referring back to \eref{Hamiltonian1} we undo the rescaling of \eref{scH}, and identify the temperature unit with the one in a BEC, and the clustering critical temperature reads
\bea{
\label{criticaltemperature}
T_c &=-\frac{4\pi n_0 }{(j_{1,1})^2}\frac{E_0}{k_B}N_\pm\simeq- 0.272 \frac{\pi \hbar^2\rho N}{m_a k_B}.
}
The clear role of the boundary condition distinguishes the clustering transition from  supercondensation, and uniquely determines the numerical factor. 
\par
As a consistency check, 
the critical temperature \eqref{criticaltemperature} can also be found by varying the entropy \eqref{entropy} along the mode $l=1$:
\bea{
	\delta S&=S-S_0 \nn\\
	&=- \left(n_0 \beta^2_c \int d^2 \mathbf{r}\; \psi^2_1 \right) \delta E + o(\delta E),
}
where $S_0=- 2\int d^2 \mathbf{r}\; (n_0/2) \ln (n_0/2)$ is the entropy of the uniform phase, and in the last step we used ~\eref{eq:newconstr-1-3}.

The energy change along the mode $l=1$ is
\bea{\label{E2}
	\delta E ={}& \frac{1}{2} \int d^2 \mathbf{r}\; \delta \sigma \delta \psi \nn\\
	\simeq{}&- \frac{1}{2} \epsilon^2 n_0 \beta_c \int d^2 \mathbf{r}\;  \psi_1^2 =\frac{1}{2}\epsilon^2,
}
where the normalization condition Eq.~\eqref{normalization} is used. 

We then immediately have
\bea{ 
	\frac{\partial S}{\partial E}\bigg|_{E=0}=- n_0 \beta^2_c \int d^2 \mathbf{r}\;  \psi^2_1=\beta_c,
}
consistent with the foregoing analysis.

In the dipole phase, 
the macroscopic dipole moment 
\bea{D\equiv|\mathbf{D}|\equiv \left|\sum \kappa_i \mathbf{r}_i\right| = \left|\int d^2 \mathbf{r}\; \mathbf{r} \sigma \right|}
serves as the order parameter, which becomes non-zero spontaneously as the system crosses the transition point. Importantly here $D$ does not refer to local order or off-diagonal 
long-range order but rather describes the global coherent structure of the clustered phase.
Near the transition, $2 \delta E=\epsilon^2$, and hence 
\bea{
D &= D_0 (E - E_c)^{\nu} + \mathcal{O}(\delta E)=  \tilde{D}_0 |\beta-\beta_c|^{\nu}+ \mathcal{O}(\delta \beta)\label{dip}
}
with the critical exponent 
\bea{\label{nuc}
\nu&=1/2, 
}
and $E_c=0$. The critical exponent $\nu$ takes the same value as the critical exponent for the condensation field in the mean field XY model. 
The coefficients are $D_0=\sqrt{- J_2(j_{1,1})/J_0(j_{1,1})}=1$ and $\tilde{D}_0=C_c/\beta_c^2$, where
\bea{\label{spE}
C_c& \equiv-\beta_c^2\frac{\partial E }{\partial \beta}\Bigg|_{\beta=\beta_c} \simeq 8.91 
} 
is the microcanonical specific heat at the vortex clustering transition point (see Appendix~\ref{appendixa}). The positiveness of the microcanonical specific heat in the neutral dipole phase suggests that ensemble equivalence may hold for the confined neutral vortex system. This behavior is in stark contrast with the single charge dipole phase where the microcanonical specific heat is negative~\cite{Smith:1990cx}.
Note that $D_0$ and $C_c$ do not depend on the chosen normalization of $\psi_1$. 
\par
Clear evidence of the emergence of global order in the vortex clustering transition can be seen in real space. In the dipole phase, $|\sigma|$ is symmetric under $x\to-x$,
and $\psi(0,y)=0$. 
Dividing the disc into two regions $A_\pm=\{x\gtrless0\}$, we define the average position of net vortices in region $A_\pm$ as
\bea{\label{center}
	\overline{\mathbf{r}}_\pm &\equiv \frac{ \int_{A_\pm} d^2\mathbf{r}\; \mathbf{r} |\sigma|} {\int_{A_\pm} d^2\mathbf{r}\; |\sigma|}. } 

The separation between vorticity centres can be measured by 
\bea{\label{dlims}
d&=|\mathbf{d}| \equiv |\overline{\mathbf{r}}_+-\overline{\mathbf{r}}_-|.}
Evaluating Eq.~\eqref{center} close to the clustering transition gives 
\bea{
d=d_c + \mathcal{O} (\delta E)} 
with 
\bea{\label{dclim}
d_c& = H^{-1}_1(j_{1,1})\simeq 0.92,
}
and $H_1(x)$ is the first-order Struve function~\cite{Grad}.

\subsection{Supercondensation}
Far above the vortex clustering transition each cluster shrinks into a rather small region centred at $r_{\pm}=|\mathbf{r}_{\pm}|\equiv|\mathbf{r}\mp\mathbf{d}/2|$. In the supercondensation regime the angular variation of the vortex distributions in the neighbourhood of $\mathbf{r}_{\pm}=0$ may be safely neglected, giving the asymptotically exact vortex density distributions~\cite{Smith:1990cx}, 
\bea{ \label{exactdensity}
	n_{\pm} =\frac{4A'}{(2-\pi \beta A' r^2_{\pm})^2},
}
with the smoothness condition $n'_{\pm}(\mathbf{r}_{\pm}=0)=0$,
where $A'$ is fixed by the 
normalization condition. For high energy, $A' \rightarrow A(\beta)$ with
\bea{\label{Aprime}
A(\beta)&\equiv\frac{1}{\pi(1-\beta/\beta_s)},
}
and $\beta \rightarrow \beta_s=-2$. The inverse temperature $\beta_s$ defines the universal supercondensation point, at which vortices concentrate spatially into point-like clusters, irrespective of either the boundary condition or the net vortex charge~\cite{Smith:1990cx}. In BEC units, supercondensation occurs at $T_s= -\pi \hbar^2\rho N/4m_a k_B$, consistent with an estimate based upon the free energy argument~\cite{KRAICHNAN:1980uy} and with canonical MC sampling~\cite{Simula:2014ku}. Viewing the clusters as points, the equilibrium condition for the two point vortices with opposite circulation yields $d\to d_s$, where
\bea{\label{dslim}
d_s&=2(\sqrt{5}-2)^{1/2}\simeq 0.972
} 
is independent of the cluster charges. This equilibrium configuration balances the force of interaction between the two clusters with the force arising interaction with the container; the latter is generated by the images in \eref{Hamiltonian1}.
The limiting forms \eref{dclim} and \eref{dslim} give the cluster separation for energies near and well above the transition, and clearly the location of the clusters is only rather weakly dependent on system energy. 

The main contribution to the energy comes from vortices with the same circulation, hence
\bea{E\simeq - 4\beta^{-2}\left[\ln\left(1-\beta/\beta_s\right)- \beta/2  \right].} 
As $\beta \rightarrow \beta_s$,
$|\langle D \rangle - \langle D \rangle_{max}|\sim \beta -\beta_s$ with $\langle D \rangle_{max}=d_s$, and the microcanonical specific heat has asymptotic form
\bea{C_s=-\beta^2 (\partial E/ \partial \beta) \propto (\beta-\beta_s)^{-1}>0.\label{specificheat}} 
The power law divergence of the microcanonical specific heat is determined by 
the asymptotically exact vortex density distribution Eq.~\eqref{exactdensity}. In any physical realization of the point-vortex model short-range repulsion becomes important in the supercondensation limit. Hence, instead of collapsing to a point associated with infinite energy, a tight crystal lattice structure would occur~\cite{Simula:2014ku}.

\section{Monte Carlo Simulations}
Starting from the uniform phase at minimum energy $E=0$, we perform Monte Carlo simulations for range of increasing energies, at fixed angular momentum $L=0$.  
\subsection{Microcanonical Monte Carlo}\label{appendixb}

Our microcanonical Monte Carlo simulations follow the general scheme of
Refs.~\cite{Smith:1989a,Smith:1990cx}, based on the Demon algorithm of Creutz \cite{Creutz:1983bo}, modified to efficiently sample the phase space of the neutral point-vortex system. Specifically, we couple the system of point vortices to a \textit{demon} with
energy and angular momentum degrees of freedom $E_\mathrm{D}$ and
$L_\mathrm{D}$, and write the total energy and angular momentum as
\begin{align}
E_\mathrm{Total} &= E + E_\mathrm{D}\,,\\
L_\mathrm{Total} &= L + L_\mathrm{D}\,,
\end{align}
where $E$ and $L$ are the point-vortex energy and angular momentum.
\begin{figure}[!t]
	\includegraphics[width=\columnwidth]{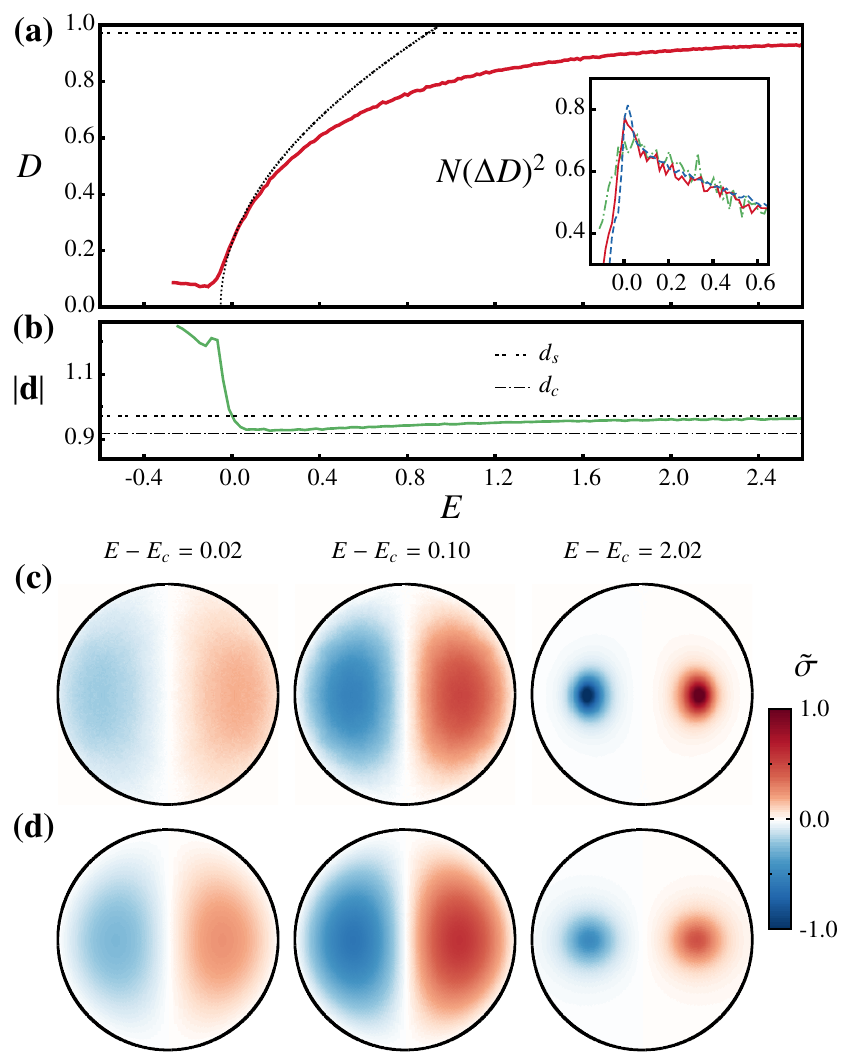}
	\caption{(Color online) A comparison between microcanonical theory and MC sampling for $N=100$ point-vortices on the clustering transition. (a) shows the
dipole moment $D$ (solid red) for increasing
energy $E$; also shown is the mean field prediction $D=(E-E_c)^{1/2}$ (dotted black). $E_c\simeq -5\times 10^{-2}$ --- here representing the critical energy for $N=100$
--- is estimated by fitting to the numerical data [see Sec.~\ref{sec:N}]. The inset in (a) shows the variance of dipole moment for $N=50$ (dash dotted green), $N=100$ (solid red), and $N=200$ (dashed blue). (b) shows
the separation between vorticity centres $d(E)$, \eref{dlims}; the numerical data (solid
line) is shown alongside asymptotic mean field predictions for high energies $d_s$~[double-dashed line, \eref{dslim}] and energies close to the transition $d_c$ [dot-dashed line, \eref{dclim}].
(c) and (d) show, respectively, results from MC-sampling and mean field theory for the
scaled vortex density $\tilde{\sigma}$ (see text). To present the large energy range on one color map, we define $\tilde{\sigma}
\equiv\sigma$ for $E-E_c = 0.02, 0.1$, and $\tilde{\sigma} \equiv \sigma/10$ for $E-E_c=2.02$.
} \label{dipole}
\end{figure}
To sample the microcanonical ensemble for desired energy and angular momentum
$E'$ and $L'$, we fix $E_\mathrm{Total} = E'$ and $L_\mathrm{Total} = L'$ and
perform a random walk subject to the constraints $|E_\mathrm{D}| <
E_\mathrm{max}$ and $|L_\mathrm{D}| < L_\mathrm{max}$. Typical constraints on
the Demon used here are $E_\mathrm{max} = 2.5\times10^{-2}$ and $L_\mathrm{max}
= 1\times10^{-2}$.

Our random walk consists of two alternating types of Monte Carlo step.  The
first Monte Carlo step is comprised of $N$ pairwise random moves of two
randomly-chosen vortices, such that $\mathbf{r}_i \rightarrow \mathbf{r}_i +
\mathbf{\delta r}_i$, where $i=1,2$ indexes the vortices, and the random moves
$\mathbf{\delta r}_i$ are chosen uniformly randomly from a square of side length
$\Delta$ centered on the origin. We update $\Delta$ to maintain an efficient
success rate for the random walk (typically such that 50\% of pairwise moves
succeed). This Monte Carlo step is very similar to that used in
Ref.~\cite{Smith:1990cx} for \textit{single-circulation} point-vortex systems.

To more efficiently explore the possible values of the dipole moment in the
microcanonical ensemble of \textit{neutral} point-vortex systems, we follow
each Monte Carlo step of the pariwise type described above with a single
Monte Carlo step of a second type.  This second Monte Carlo step consists of a
similar pairwise move, but in this case applied to a pair consisting of; (a)
the center of circulation of all positive vortices, and (b) the center of
circulation of all negative vortices. Such a move is achieved by moving
\textit{all} the positive (negative) vortices by a vector $\mathbf{\delta r}_+$
($\mathbf{\delta r}_-$), such that
\begin{align}
\frac{1}{N_+}\sum_{\kappa_i >0} \mathbf{r}_i &\rightarrow \frac{1}{N_+}\sum_{\kappa_i >0} \mathbf{r}_i + \mathbf{\delta r}_+,\\
\frac{1}{N_-}\sum_{\kappa_i <0} \mathbf{r}_i &\rightarrow \frac{1}{N_-}\sum_{\kappa_i <0} \mathbf{r}_i + \mathbf{\delta r}_-.
\end{align}
The random moves $\mathbf{\delta r}_\pm$ are chosen from the same distribution
as the pairwise moves above, using the same value of $\Delta$. The addition of this second type of step leads to greater
ergodicity in the values of dipole moment obtained during the random walk.

\renewcommand{\tabcolsep}{12pt}
\begin{table}[!t]
	\centering 
	\begin{tabular}{ccc}
	\hline\hline
		$N$ & $E_c\;[ 10^{-2}]$ & $\Delta E_c \;[ 10^{-2}]$\\
		\hline
		$50$ & $-9$ & $3$\\
		$100$ & $-5$ & $1$\\
		$200$ & $-2$ & $1$\\
		$500$ & $-1.0$ &  $0.7$\\
		$1000$ & $0$ &$0.2$\\
		\hline\hline
	\end{tabular}
	\caption{Estimated values of $E_c(N)$ and its uncertainty $\Delta E_c(N)$ obtained by our numerical fitting procedure [see Sec.~\ref{sec:N}].\label{table:e_c}}
\end{table}
\subsection{Vortex Clustering Transition for $N=100$}
A comparison between the analytical predictions of mean field theory and the microcanonical MC sampling for $N=100$ point-vortices is shown in Fig.~\ref{dipole}. The numerical results for the values of $\nu$ and $D_0$ are in good agreement with the mean field predictions. Before the transition $\langle D \rangle$ has a small value which is almost independent of energy, due to the finite value of $N$. Here the brackets refer to the ensemble average. The variance of the order parameter $(\Delta D)^2 \equiv \langle D^2 \rangle - \langle D \rangle^2 $ is evaluated numerically 
and it shows a peak around the transition point, where the dipole structure becomes unstable (see Fig.~\ref{dipole}(a) inset). Moreover, $N (\Delta D)^2$ grows with increasing $N$ near the transition point while it has almost no $N$-dependence elsewhere.  These non-Gaussian fluctuations of the dipole moment indicate non-trivial correlations between vortices due to criticality.	
As a check of our MC-sampling, we 
independently verified the results in Fig.~\ref{dipole} (a) by sampling using the Wang-Landau algorithm \cite{Wang:2001eb}.
\par
The values of $d_c$ and $d_s$ are also confirmed by MC simulations as shown in Fig.~\ref{dipole}(b).
Note that $d\neq0$ before the transition in our MC simulations, again due to the finite value of $N$. In crossing the transition $\overline{\mathbf{r}}_\pm$
jumps from zero to a nearly energy-independent constant.
Positive (negative) vortices accumulate at the point  $\overline{\mathbf{r}}_\pm$ gradually as energy increases, which yields continuous growth of the dipole moment $\langle D \rangle$. 
\par
In Fig.~\ref{dipole}, (c) and (d) we compare MC-sampling and mean field theory for the density difference $\sigma$ at different energies. For $E-E_c = 0.02, 0.1$, the mean field result for $\sigma$ is evaluated perturbatively via Eq.~\eqref{selfconsistenteq} by plugging in the zero mode Eq.~\eqref{dipMode}. 
Agreement between the spatial distributions is very good near the clustering transition where a single unstable mode is dominant. At the high energy $E=2.02$, we compare MC-sampling with the supercondensation asympotic form [\eref{exactdensity}] that ignores the interaction between clusters. The MC-sampling reveals additional elliptical distortion of each cluster.
\subsection{Order Parameter Dependence on $E$ for Increasing $N$}\label{sec:N}
We also extract the
average dipole moment $D$ as a function of energy for a range of $N$. 
\begin{figure}[!t]
	\centering
	\includegraphics[width=\columnwidth]{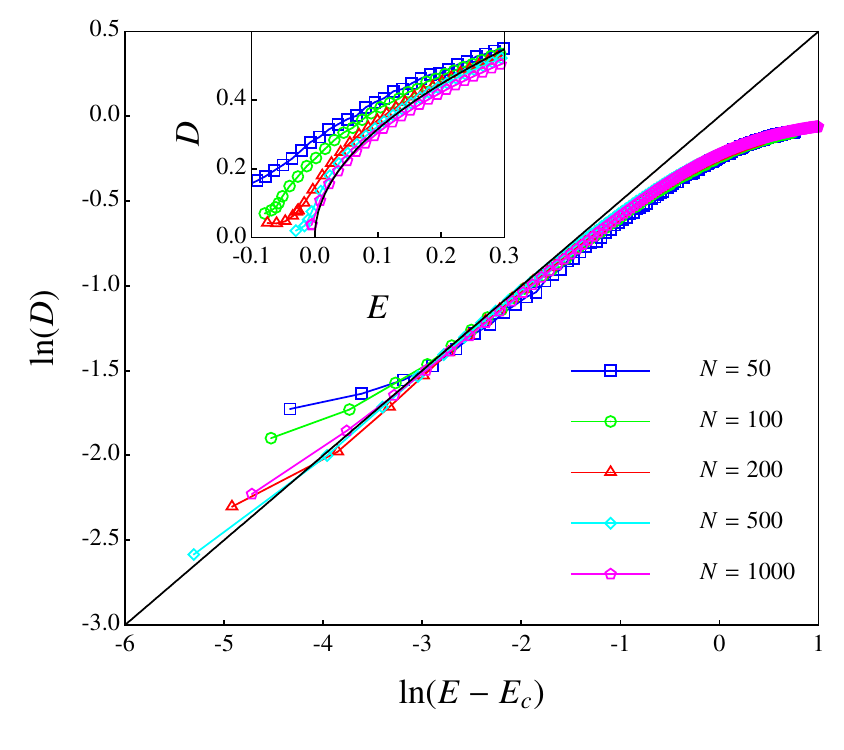}
	\caption{(Color online) Logarithmic plot of average dipole moment $D$ against the energy above the critical energy, $E-E_c$. Solid line shows the predicted mean-field scaling $|E-E_c|^{1/2}$. Inset shows a non-logarithmic plot of $D$ against $E$.\label{fig:logplot}}
\end{figure}
We first estimate the critical energy $E_c$ for each $N$ using a simple fitting procedure in the region close to the transition. Given discrete samples of the dipole moment $D_i=D(E_i)$, and its variance, obtained from MC simulations conducted at energies $E_i$, we minimize
\bea{\label{chi2}
\chi^2& = \sum_{i \in \mathcal{R}} \frac{\left( D_i - \sqrt{E_i-E_c} \right)^2}{(\Delta D_i)^2}\,,
}
where $\mathcal{R}$ is the set of $i$ such that $0 < E_i-E_c < 0.1$. This choice of energy range leads to at least 7 points in the set $\mathcal{R}$ for every value of $N$. This procedure yields the values of $E_c$ reported in Table~\ref{table:e_c}.  We can see that $E_c(N)\to 0 $, the mean field prediction, as $N$ increases.  We calculate an estimate of the uncertainty in $E_c(N)$ from the $\chi^2+1$ contour, $\Delta E_c(N)$. The error in $E_c$ decreases with increasing $N$, corresponding to increasing goodness of fit~\cite{Hughes:2010vd}.

Having estimated $E_c$ independently for each $N$, in Fig.~\ref{fig:logplot} we plot dipole moment $D$ against $E-E_c$ on a log scale. The scaling in the transition region becomes very close to $|E-E_c|^{1/2}$ as $N$ increases, with good agreement for the scaling already evident for $N=100$. Note that the finite values
of $D$ observed below the transition energy systematically decrease in
amplitude with increasing $N$ [Fig.~\ref{fig:logplot} (inset)]. Also note that
the lowest accessible energy of the finite-size system increases with $N$; in the
thermodynamic limit there are no states with $E<0$.

\section{Discussion and conclusions} 
\subsection{Phases of the (Extended) Point-Vortex Model}
\begin{figure}[!t]
	\begin{center}
		\includegraphics[width=\columnwidth]{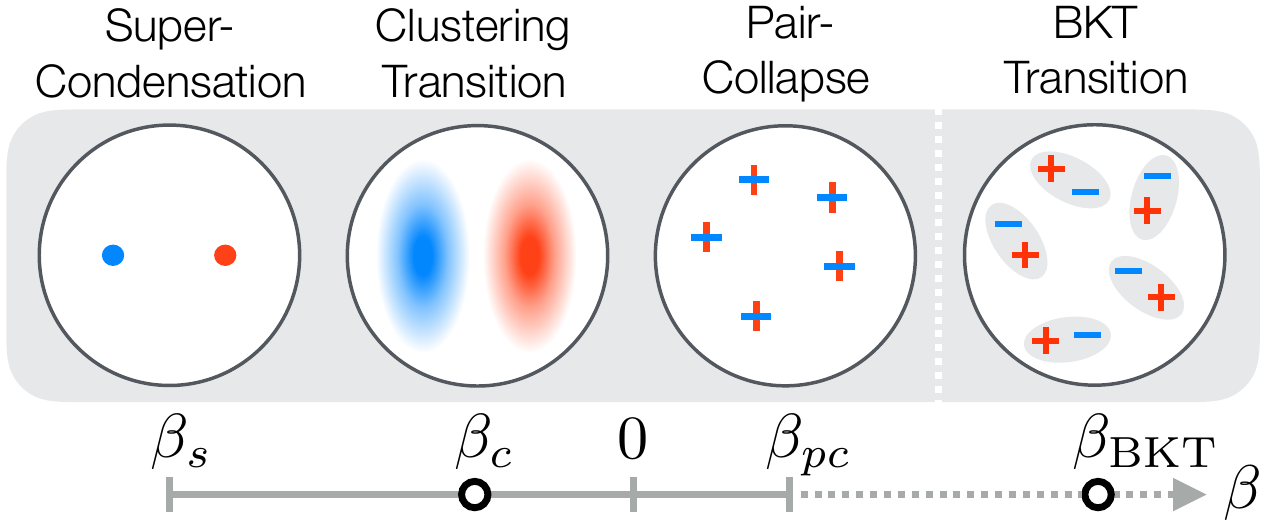}
		\caption{(Color online) Schematic of phases of a neutral system of point vortices confined to the disc geometry. Energy per vortex decreases from left to right. In dimensionless units the supercondensation temperature occurs at the universal point $\beta_s=-2$. The vortex clustering transition occurs at $\beta_c\simeq-1.835$ [\eref{bc}]. 
		In the positive temperature regime, the pair-collapse limit is reached at $\beta_{pc}$. The addition of short-range repulsion between vortices allows the point-vortex model to extend further into the positive temperature regime to encompass the BKT transition at $\beta_{\rm BKT}>\beta_c$ (see text).
\label{schematic}} 
	\end{center}
\end{figure}

In Fig.~\ref{schematic} we summarize the equilibrium phases of the system of point-vortices confined to a disc. As an end
point of a continuous process as energy increases, the system enters the supercondensation regime at $\beta_s=-2$~\cite{Kraichnan:1975ku}, involving point-like concentration of vortices with the same circulation, and divergence of the system energy. At $\beta_c$, with $\beta_s<\beta_c<0$, a clustering transition occurs, whereby large-scale aggregation of like-sign vortices can occur, breaking the underlying symmetries of the homogeneous phase.  
Our microcanonical ensemble approach demonstrates that the clustering transition is strongly dependent on the geometry of the system, while supercondensation is universal. Note that in the canonical ensemble where states are parametrized by inverse temperature, the clustering transition and supercondensation are difficult to distinguish~\cite{Simula:2014ku}. The microcanonical approach reveals that the apparent coincidence of the transitions is due to numerical proximity of $\beta_s$ and $\beta_c$. In the microcanonical ensemble, states are parametrized by energy, giving a clear distinction between the vortex clustering transition and supercondensation.
\par
Our mean field analysis gives analytical expressions for the structure of the clustered phase and critical behavior of the dipole moment near $\beta \simeq \beta_c$. 
For $\beta_c<\beta<0$, Eqs.~\eqref{Poisson}, ~\eqref{selfconsistenteq} only have the trivial solution, namely $n_{\pm}(\mathbf{r})=n_0$ and $\psi=0$, which gives the limitation of our current mean field approach: the states at $\beta_c<\beta<0$ are not distinguishable. In addition, our mean field approach is not suitable to investigate the $\beta>0$ phases, instead the proper continuous effective field formulation of the point-vortex model is the Sine-Gordon field theory. At $\beta_{pc}>0$ the model is known to exhibit a collapse of oppositely charged pairs~\cite{Hauge:1971tt,Cornu:1987ih,Cornu:1989dl}. Going beyond the pure point-vortex model, introducing a short-range repulsion between vortices associated with a finite vortex core scale softens the pair-collapse into a Berezinsky$-$Kosterlitz$-$Thouless (BKT) transition at $\beta_{\rm BKT}$~\cite{Kosterlitz:1973fc,Kadanoff:1978gc,Nienhuis:1984dj}. 
\subsection{Symmetries and Order Parameter}
One should distinguish the vortex clustering transition discussed in this paper from the BEC transition. The fundamental degrees of freedom in our system are the vortices but not the bosonic atoms of the host fluid. The clustering transition is not described by a \emph {local} order parameter. The macroscopic dipole moment instead describes the global spatial structure of the system. The clustered phase breaks the underlying $\rm SO(2)$ symmetry as well as the $\mathbb Z_2$ symmetry, and we emphasize that the $\rm SO(2)$ symmetry breaking occurs at zero angular momentum. There are also clustered phases which break the $\mathbb Z_2$ symmetry while preserving the $\rm SO(2)$ symmetry, carrying finite angular momentum. The macroscopic dipole moment $\left<D\right>=\left< |\sum_{i}\kappa_i {\mathbf r}_{i}| \right>$ functions as a global order parameter as $\left<D\right>$ provides a good description of the transition from the uniform phase $\left<D\right>=0$ to the clustered dipole phase $\left<D\right> \neq 0$.  
The critical exponent for the order parameter, \eref{nuc}, has the same value as the mean field XY model. However an anomalous scaling correction might be present due to a remnant of some screening effects close to the transition, which is beyond the scope of the mean field theory approach taken here.  Near the transition we expect that $N(\Delta D)^2\sim (E-E_c)^{-\gamma}$ as $N\to\infty$. While our current level of MC sampling is sufficient to determine the scaling of $\langle D\rangle$, identifying $\gamma$ would require significant further investigation.

	The clustering transition is not described by a local order parameter and hence one can not apply the standard scenario of phase transitions straightforwardly. However, the clustering transition does have features of a second-order phase transition: 1) it spontaneously breaks the underlying symmetries; 2) there is an order parameter associated with the broken symmetry that grows continuously as a function of the tuning parameter crossing the transition; 3) the fluctuation of the order parameter has a peak near the transition and the peak grows as the number of vortices $N$ increases.       
\par
The important role of symmetry in this system warrants some discussion of the role of the SO(2) symmetry of the trap. 
We note that a very high degree of cylindrical trap symmetry was essential for preserving sufficient angular momentum to achieve rotating Bose-Einstein condensation~\cite{Haljan2001}. Furthermore, recent advances in digital-micromirror device technology~\cite{Gauthier:2016vs} could allow the construction of highly circular traps. Another promising avenue may be offered by using specific Gauss-Laguerre modes to induce radial confinement~\cite{Ramanathan:2011bi}, as laser fields can be efficiently shaped into specific modes that are SO(2) symmetric.
\subsection{Conclusions}
We study clustering phenomena in the neutral point-vortex model confined to a disc, in the microcanonical ensemble. Utilizing concepts from the standard theory of phase transitions, we give a clear picture of the clustering transition of quantum vortices, including the transition temperature, the power law scaling of the order parameter near the transition, and the distinction between clustering and supercondensation. The clustering transition breaks the underlying $\mathbb Z_2$ and $\rm SO(2)$ symmetries, exhibiting features of a mean field $\rm XY$-type second-order phase transition. 
The mean field theory is found to agree well with microcanonical Monte Carlo simulations, consistent with the dominance of long-range interactions in the clustered regime. 
By uncovering the nature of the clustering transition in a bounded domain, our theory gives a treatment of the transition relevant for experimental realization in Bose-Einstein condensates, and suggests future directions for testing ensemble equivalence in systems with long-range interactions.
		Our approach could be extended to arbitrary confining geometries, and provides a starting point for studying  vortex clustering phenomena in quantum fluid systems that are strongly forced and damped~\cite{Lagoudakis08a,Harris:2016hl}. 
		
\section*{Acknowledgments} 
We acknowledge M. M\"{u}ller, P. B. Blakie,  T. P. Simula, S. A. Gardiner, S. Flach, S. Denisov, T. Li and A. M. Mateo for useful discussions. 
TPB acknowledges support from the John Templeton Foundation via the Durham Emergence Project.
ASB is supported by a Rutherford Discovery Fellowship administered by the Royal Society of New Zealand. 

\appendix

\section{Specific Heat Near the Transition} \label{appendixa}
In order to investigate the microcanonical specific heat at the transition point,
we need to study higher-order perturbation theory in $\epsilon$.
Similar analyses for single charge vortices can be found in Ref.~\cite{Smith:1990cx}. In the following we always use the fact that at the transition point $\sigma=0$ and the rotation-less condition $\alpha=\delta \alpha=0$.

Let us first consider higher-order corrections of $\delta \psi$ for $l=1$,
$\delta \psi \simeq \epsilon \psi_1 + \psi^{(2)}_1 + \psi^{(3)}_1$, where $\psi^{(i)}$ $(i = 2, 3)$ 
is of order $\mathcal{O} (\epsilon^i)$.   
To the $\mathcal{O} (\epsilon^2)$, 
\bea{
	\label{variationPoisson}
	\nabla^2 \delta \psi ={}&-4 \pi \left[ \delta n_+ (\mathbf{r}) - \delta n_- (\mathbf{r})\right] \nn\\
	={}&4 \pi \left[n_0 (\psi \delta \beta + \beta \delta \psi + r^2 \delta \alpha)- n_+ \delta \gamma_+ +  n_- \delta \gamma_- \right] 
}
becomes
\bea{
	\label{Poisson-o(2)-d}
	{\cal L}_{0} \psi^{(2)}_1 &= 4 \pi \bigg(-n_+  \delta \gamma_+ + n_- \delta \gamma_- \bigg),
}
where ${\cal L}_{0}\equiv\nabla^2-4\pi n_0\beta$.
Multiplying $\psi_1 (r, \theta) = \hat{\psi}_1 (r) \cos(\theta)$ 
to \eref{Poisson-o(2)-d} and integrating over space, 
the right-hand side of the equation is trivially zero and the left-hand side of the equation is also zero because   
$\int d^2 \mathbf{r}\; {\cal L}_0 \psi^{(2)} \psi_1=\int d^2 \mathbf{r}\; \psi^{(2)} {\cal L}_0 \psi_1=0$.
We introduce the function $\eta$ which satisfies 
\bea{
	\label{eta}
	\left(\frac{1}{r} \frac{d}{dr} r \frac{d}{dr} - 4 \pi n_0 \beta_c \right) \eta  &= 2 \pi n_0   
}
with boundary conditions $\eta (r = 1) = 0 $ and $\eta'(r=0)=0$.
The solution of \eref{eta} reads $\eta (r)=[2\beta_c J_0(j_{1,1})]^{-1} J_0(j_{1,1} r) - 1/2\beta_c$. Then $\psi^{(2)}_1$ can be expressed as 
\bea{
	\psi^{(2)}_1&=\eta (-\delta \gamma_+ + \delta \gamma_-).
}
Conservation of the number of vortices leads to 
\begin{align}\label{eq:newconstr-1}
\int d^2 \mathbf{r}\; (\delta n_+ - \delta n_-) &= 0,\\
\int d^2 \mathbf{r}\; (\delta n_+ + \delta n_-)&= 0\, .\label{eq:newconstr-3}
\end{align}
At $\mathcal{O} (\epsilon^2)$, the constraint \eref{eq:newconstr-1} becomes
\bea{
	\int d^2 \mathbf{r}\;  \nabla^2 \delta \psi &= \int d^2 \mathbf{r}\; \nabla^2 \psi^{(2)}_1=0,
}
and then we have
\bea{
	\int_0^{2\pi} d\theta \frac{\partial \psi^{(2)}_1}{\partial r}\bigg|_{r=1}&=0, \label{eq:newconstr-1-2}
}
which is satisfied automatically. 
Then \eref{eq:newconstr-3} takes the following form
\bea{
	\label{eq:newconstr-1-3}
	\int d^2 \mathbf{r}\; \bigg(n_+\delta \gamma_++n_-\delta \gamma_- +\frac{1}{2}n_0\beta^2 \epsilon^2 \psi_1^2 \bigg)&=0.
}
At $\mathcal{O} (\epsilon^2)$, the constraint $0=\delta M=-\pi^{-1}\int d^2\mathbf{r}\; \psi^{(2)}$ requires that $\delta \gamma_-=\delta \gamma_+$, and therefore $\psi^{(2)}_1=0$.

In order to obtain the $\beta-E$ relation, we need to expand \eref{variationPoisson} to $\mathcal{O} (\epsilon^3)$:
\bea{
	\label{Poisson-o(3)-d}
	{\cal L}_0 \psi^{(3)}_1&=2 \pi n_0 \epsilon \psi_1 \bigg[ 2\delta \beta 
	+ \beta_c\delta \gamma_+  + \beta_c \delta \gamma_- + \frac{\epsilon^2}{3}\beta^{3}_c \psi_1^2 \bigg].
}
Multiplying $\psi_1 (r, \theta)$
to \eref{Poisson-o(3)-d} and integrating over space, 
the left-hand side of the equation is zero because   
$\int d^2 \mathbf{r}\; {\cal L}_0 \psi^{(3)} \psi_1=\int d^2 \mathbf{r} \;\psi^{(3)} {\cal L}_0 \psi_1=0$. Hence we have
\bea{
	\label{Fredhomsolubility}
	\epsilon  \int d^2 \mathbf{r}\; \psi_1^2\left[2\delta \beta + \beta_c (\delta \gamma_+  + \delta \gamma_-) \right]+\frac{\epsilon^3}{3} \beta^3_c \int d^2 \mathbf{r}\;  \psi_1^4=0. 
}
Combining \eref{eq:newconstr-1-3} and \eref{Fredhomsolubility}, we obtain $\delta \beta = \tilde{C}\delta E$, where 
\bea{
	\tilde{C}={}& \beta^3_c\bigg[\frac{n_0}{2} \bigg(\int d^2 \mathbf{r}\; \psi^2_1 \bigg)^2 -\frac{1}{3} \int d^2 \mathbf{r}\; \psi^4_1 \bigg] \bigg( \int d^2 \mathbf{r}\; \psi^2_1 \bigg)^{-1}  \nn\\
	\simeq{}& -0.378.
}
Hence we arrive at \eref{spE}.

%

\end{document}